\documentclass[aps,prd,10pt,notitlepage,nofootinbib]{revtex4-1}

\usepackage[utf8]{inputenc}
\usepackage{amsmath,amssymb,amsfonts}
\usepackage{newtxtext,newtxmath}

\usepackage{bbold}
\usepackage{bm}
\usepackage{graphicx}
\usepackage[usenames,dvipsnames]{xcolor}
\usepackage{color}
\usepackage[colorlinks=true,linkcolor=blue,urlcolor=blue,citecolor=blue]{hyperref}

\usepackage{slashed}
\usepackage[english]{babel}
\usepackage{dcolumn}
\usepackage{pifont}
\usepackage{dsfont,mathrsfs}
\usepackage{cancel}
\usepackage{bigints}
\usepackage{accents}
\usepackage{soul}
\usepackage{multirow}

\usepackage{amsmath, amssymb}
\usepackage{bbold}
\usepackage{makecell}
\usepackage{simpler-wick}
\usepackage{setspace}
\usepackage{scalerel}
\usepackage{orcidlink} 
\usepackage[normalem]{ulem}
\usepackage{hyperref}

\newcommand{\nn}{\nonumber}

\newcommand{\FB}[1]{\left(#1\right)}
\newcommand{\fb}[1]{(#1)}

\newcommand{\TB}[1]{\left[#1\right]}

\newcommand{\munu}{{\mu\nu}}

\newcommand{\IM}{\text{Im}}

\newcommand{\util}{\widetilde{u}}
\newcommand{\btil}{\widetilde{b}}
\newcommand{\Ncal}{\mathcal{N}}
\newcommand{\Pcal}{\mathcal{P}}

\begin{document}
	\title{Ellipticity of dilepton production from a hot and magnetized hadronic medium}

	\author{Rajkumar Mondal\orcidlink{0000-0002-1446-6560}$^{a,d}$}
	\email{rajkumarmondal.phy@gmail.com}

	\author{Nilanjan Chaudhuri\orcidlink{0000-0002-7776-3503}$^{a,d}$}
	\email{sovon.nilanjan@gmail.com}
	\email{n.chaudhri@vecc.gov.in}
	
	\author{Snigdha Ghosh\orcidlink{0000-0002-2496-2007}$^{b}$}
	\email{snigdha.ghosh@bangla.gov.in}
	\email{snigdha.physics@gmail.com}
	
	\thanks{Corresponding Author}
	
	\author{Sourav Sarkar\orcidlink{0000-0002-2952-3767}$^{a,d}$}
	\email{sourav@vecc.gov.in}
	
	\author{Pradip Roy$^{c,d}$}
	\email{pradipk.roy@saha.ac.in}

	\affiliation{$^a$Variable Energy Cyclotron Centre, 1/AF Bidhannagar, Kolkata - 700064, India}
	\affiliation{$^b$Government General Degree College Kharagpur-II, Paschim Medinipur - 721149, West Bengal, India}
	\affiliation{$^c$Saha Institute of Nuclear Physics, 1/AF Bidhannagar, Kolkata - 700064, India}
	\affiliation{$^d$Homi Bhabha National Institute, Training School Complex, Anushaktinagar, Mumbai - 400085, India} 
	

\begin{abstract}
	We study the azimuthal angle and transverse momentum dependence of dilepton production from hot and magnetized hadronic matter using $\rho^0$-meson dominance. The thermomagnetic spectral function of the $\rho^0$ is evaluated using the real time method of thermal field theory and Schwinger proper-time formulation. A continuous spectrum is obtained in which there is sizeable Landau cut contributions in the low invariant mass region as a consequence of finite background field. The emission rate of the dileptons is found to be significantly anisotropic in this region and the later effectively increases with the strength of the background field. In addition, we also evaluate the elliptic flow parameter ($v_2$) as a function of invariant mass for different values of magnetic field and temperature. We find that in low invariant mass region $v_2$ remains positive at lower values of $eB$ signifying that the production rate could be larger along the direction transverse to the background field. This behaviour is consistent with the angular dependence of dilepton production rate.
\end{abstract}

\maketitle

\section{Introduction}
It is well established that deconfined nuclear matter is created in highly energetic heavy ion collision (HIC) experiments at Relativistic Heavy Ion Collider (RHIC) at Brookhaven and Large Hadron Collider (LHC) at CERN. This novel state of matter is known as quark-gluon-plasma (QGP) which is a deconfined state of quarks and gluons in local thermal equilibrium~\cite{Wong:1995jf,Sarkar:2010zza,Florkowski:2010zz,Satz:2012zza}. The primary objective of the ongoing HIC experiments is to study the nature as well as the fundamental properties of this strongly interacting system. Because of the transient nature of the produced matter one has to look for indirect signals of which electromagnetic probes (photons and dileptons), due to their large mean free paths, are the most efficient to characterize the space-time history of the collision~\cite{McLerran:1984ay,Kajantie:1986dh,Gale:1988vv,Weldon:1990iw,Ruuskanen:1990hx,Ruuskanen:1991au,Alam:1996fd,Alam:1999sc}. 

It is conjectured that in non-central HICs, a very strong magnetic field is  produced due to the rapid movement of the electrically charged spectators in the early stage of the collision. The estimated value of the strength of this magnetic field is $\approx 10^{15-18}$ Gauss~\cite{Kharzeev:2007jp,Skokov:2009qp} which decays very rapidly with time within a few fm/c. However, due to the finite conductivity ($\sim$few MeV) of the produced medium, decay of the magnetic field is expected to be sufficiently delayed~\cite{Tuchin:2013apa,Tuchin:2015oka,Tuchin:2013ie,Gursoy:2014aka,Inghirami:2016iru,Inghirami:2019mkc,Kalikotay:2020snc}
 so that a non-zero magnetic field persists even during the hadronic phase which is realized after a phase transition/crossover from the QGP. 
 Consequently, the properties of the hadronic matter could be significantly modified. This has led to several works in this direction of which we mention a few. 
 The effects of magnetic field on the transport properties from hadronic medium have been studied in Refs.~\cite{Kadam:2014xka,Das:2019pqd,Dash:2020vxk,Ghosh:2022xtv,Das:2019wjg,Kalikotay:2020snc,Das:2020beh}. Estimation of shear and bulk viscosity has been made in different approaches from magnetically modified hadronic matter in Refs.~\cite{Kadam:2014xka,Das:2019pqd,Dash:2020vxk,Ghosh:2022xtv}. The effects of magnetic field on the electrical conductivity from a strongly interacting hadron gas has been studied in Refs.~\cite{Das:2019wjg,Kalikotay:2020snc}. The results about the directed flow and its role in studying the initial magnetic field is explored in~\cite{Das:2016cwd,Chatterjee:2018lsx,STAR:2019clv,ALICE:2019sgg}.

The emission of electromagnetic quanta such as photons and dileptons is also expected to be modified in the presence of a background magnetic field. 
Dilepton production from QGP medium in presence of a background magnetic field has been extensively studied in the literature by many authors~\cite{Tuchin:2012mf,Tuchin:2013bda,Sadooghi:2016jyf,Bandyopadhyay:2016fyd,Bandyopadhyay:2017raf,Ghosh:2018xhh,Islam:2018sog,Ghosh:2020xwp,Hattori:2020htm,Chaudhuri:2021skc,Wang:2022jxx,Das:2021fma} using different approximations. 
%
%
The ellipticity of dilepton emission from a hot magnetized QCD plasma has been studied in Refs.~\cite{Wang:2022jxx}.
Now, as the system expands and cools from a QGP state, hadronic matter will be generated which is known to contribute significantly to dilepton production in the low invariant mass region. Recently in Ref.~\cite{Mondal:2023vzx} dilepton production from a hot and dense magnetized hadronic medium has been studied in terms of the in-medium spectral function of $\rho^0$ which is evaluated from the electromagnetic vector current correlation function using the real time formalism  of thermal field theory.
 In the present article, this study is extended to investigate the angular dependence of dilepton emission rate as well as the  ellipticity of dilepton production from a hot and magnetized hadronic medium. To the best of our information such a study has not yet been performed.

The article is organized as follows. The general formalism of anisotropy in dilepton emission is discussed in Sec.~\ref{Formalism} followed by the calculation of dilepton production rate (DPR) in a thermomagnetic hadronic medium in terms of the vector meson spectral function. The numerical results are discussed in Sec.~\ref{Numerical} and we summarize and conclude in Sec.~\ref{SC}.
\section{FORMALISM}\label{Formalism}
\subsection{Anisotropy of Dilepton Production}
\begin{figure}[h]
	\includegraphics[scale=0.5]{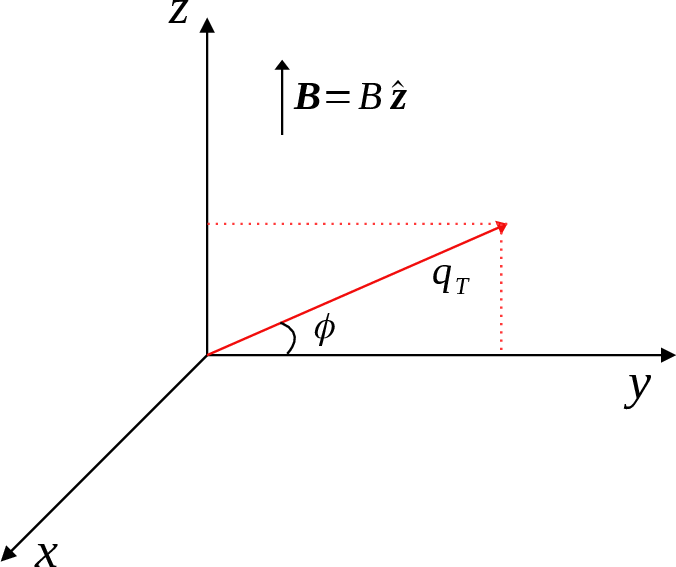}
	\caption{A schematic diagram of the coordinate frame used in this work}
	\label{Frame}
\end{figure}

To study the anisotropy in dilepton emission rate in the presence of a background magnetic field, we use the geometry shown in Fig.~\ref{Frame}. We assume that the constant external magnetic field $\bm{B}=B\hat{\bm{z}}$ points in the positive-$\hat{\bm{z}}$ direction and the beam to be along the $\hat{\bm{x}}$-direction without any loss of generality. Therefore, the transverse component of dilepton momentum $q_T$ lies in the $yz$ plane which is perpendicular to the beam direction. The azimuthal angle $\phi$ is measured with respect to the reaction plane which is the $xy$ plane in Fig.~\ref{Frame}.

The dependence of the differential dilepton emission rate in non-central HICs on the azimuthal angle relative to the reaction plane is conventionally expanded in a Fourier series as~\cite{Voloshin:1994mz},
\begin{equation}
	\frac{dN}{d^4xd^4q}=\frac{dN}{d^4xMdMq_Tdq_Tdy}\frac{1}{2\pi}\TB{1+2\sum_{n=1}^{\infty}v_n\cos\TB{n(\phi-\phi_{r})}}
\end{equation}
where $\phi_{r}$ is the angle of the reaction plane. The dilepton four momentum $q^\mu \equiv (q^0,q_x,q_y,q_z)\equiv(q^0,q_x,q_T\cos\phi,q_T\sin\phi)$ so that $q_T=\sqrt{q_y^2+q_z^2}$ is the transverse momentum, and $q_x$ is the longitudinal momentum. The infinitesimal four-momentum element $d^4q=MdMq_Tdq_Td\phi dy$, where $M=\sqrt{q^2}$ is the invariant mass and the dilepton rapidity $y=\frac{1}{2}\ln\FB{\frac{q_0+q_x}{q_0-q_x}}$ such that $q_x=0$ corresponds to central rapidity.

The flow coefficients ($v_n$) are then obtained as 
\begin{equation}
	v_n=\int_{0}^{2\pi}d\phi\cos(n\phi)\frac{dN}{d^4xd^4q} \Bigg/ \int_{0}^{2\pi}d\phi\frac{dN}{d^4xd^4q}  \label{Vn}
\end{equation}
where $v_1$ and $v_2$ correspond to the directed and elliptic flows respectively. The normalization factor in the denominator of Eq.~\eqref{Vn} is the total differential rate. The calculation of DPR=$\frac{dN}{d^4xd^4q}$ from hadronic medium at finite temperature in the presence of background magnetic field has been discussed in the next subsection.

\subsection{DPR from Magnetized Hadronic Medium at Finite Temperature}
The DPR from a thermal system of hadrons has been studied earlier by many authors~\cite{Kajantie:1986dh,Gale:1988vv, Alam:1999sc,Ruuskanen:1989tp,Mallik:2016anp}. In the notation of Ref.~\cite{Mondal:2023vzx} where the DPR from magnetized hadronic matter has recently been studied using Vector Meson Dominance (VMD) model~\cite{Gale:1990pn,Mallik:2016anp,Rapp:1999ej}, the dilepton multiplicity $dN$ per unit phase-space four-volume $d^4xd^4q$ is given by,   
\begin{equation}
	\frac{dN}{d^4xd^4q}=\frac{\alpha^2}{\pi^3q^2}f_\text{BE}(q^0)L(q^2)F_\rho^2m_\rho^2\mathcal{S}(q;T,eB)
	\label{DPR}
\end{equation}
where $\alpha$ is the fine structure constant, $F_\rho = 156$ MeV is related to the vector-meson--photon ($\rho^0-\gamma$) coupling, $L(q^2)=\FB{1+\frac{2m_l^2}{q^2}}\sqrt{1-\frac{4m_l^2}{q^2}}$, $m_l$ is mass of a lepton, $m_\rho$ is the mass of the $\rho$-meson, and $f_\text{BE}(E)=\FB{e^{E/T}-1}^{-1}$ is the Bose-Einstein thermal distribution function. In Eq.~\eqref{DPR}, $\mathcal{S}(q;T,eB)$ is the in-medium spectral function of $\rho^0$, which in turn can be expressed in terms of the imaginary part of exact/complete $\rho^0$ propagator ${D}_{\mu\nu}\FB{q;T,eB}$ in a interacting thermomagnetic background as, 
\begin{eqnarray}
	\mathcal{S}(q;T,eB) = -\frac{1}{3}g^{\mu\nu}~\text{Im}{D}_{\mu\nu}\FB{q;T,eB}. \label{eq.z.1}
\end{eqnarray}

The complete $\rho^0$ propagator ${D}_{\mu\nu}\FB{q;T,eB}$ contains all the dynamics of the hadronic medium and can be obtained by solving the Dyson-Schwinger equation in terms of thermomagnetic self energy function $\Pi^{\mu\nu}(q;T,eB)$ of $\rho^0$-meson as~\cite{Mallik:2016anp,Bellac:2011kqa}, 
\begin{equation}
	D^{\mu\nu}=D^{\mu\nu}_{0}-D^{\mu\alpha}_{0}\Pi_{\alpha\beta}D^{\beta\nu}\label{Dyson}
\end{equation} 
where $D^{\mu\nu}_{0}(q) = \left(-g^{\mu\nu}+\frac{q^\mu q^\nu}{m_\rho^2}\right)\frac{-1}{q^2-m_\rho^2+i\epsilon}$ is the bare Feynman propagator of $\rho$-meson. In order to solve Eq.~\eqref{Dyson} for $D^{\mu\nu}(q)$, it is useful to decompose the self energy $\Pi_{\munu}$ in a suitable tensor basis. As in Refs.~\cite{Ghosh:2019fet,Ghosh:2020qvg}, we use the following Lorentz decomposition of $\Pi_{\munu}$ in a thermomagnetic background: 
\begin{equation}
	\Pi^{\mu\nu}(T,eB)=\Pi_L \Pcal_L^{\mu\nu}+\Pi_A \Pcal_A^{\mu\nu}+\Pi_B \Pcal_B^{\mu\nu}+\Pi_C \Pcal_C^{\mu\nu}. \label{eq.PiB.dec}
\end{equation}
In Eq.~\eqref{eq.PiB.dec}, $\Pcal_A^{\mu\nu}$, $\Pcal_B^{\mu\nu}$, $\Pcal_C^{\mu\nu}$ and $\Pcal_L^{\mu\nu}$ are the orthogonal basis tensors given by~\cite{Ghosh:2019fet},
	\begin{eqnarray}
	\Pcal_L^{\mu\nu}~=~\frac{\util^\mu\util^\nu}{\util^2},~~\Pcal_A^{\mu\nu}~=~\FB{g^{\mu\nu}-\frac{q^\mu q^{\nu}}{q^2}-\frac{\util^\mu\util^\nu}{\util^2}-\frac{\btil^\mu\btil^\nu}{\btil^2}},~~\Pcal_B^{\mu\nu}~=~\frac{\btil^\mu\btil^\nu}{\btil^2},~~\Pcal_C^{\mu\nu}~=~\frac{1}{\sqrt{\util^2\btil^2}}\FB{\util^\mu\btil^\nu+\util^\nu\btil^\mu}\label{eq.Q}
\end{eqnarray}
where, $\util^\mu=u^\mu-\FB{\frac{q\cdot u}{q^2}}q^\mu$, $\btil^\mu = b^\mu-\FB{\frac{q\cdot b}{q^2}}q^\mu-\FB{\frac{b\cdot\util}{\util^2}}\util^\mu$, $u^\mu$ is medium four-velocity, $b^\mu = \frac{1}{2B}\varepsilon^{\mu\nu\alpha\beta} F^\text{ext}_{\nu\alpha}u_\beta$, and $F^\text{ext}_{\nu\alpha}$ is the electromagnetic field strength tensor corresponding to the external magnetic field. It may be noted that, in local rest frame (LRF) of the medium, $u^\mu_\text{LRF}\equiv(1,\bm{0})$ and $b^\mu_\text{LRF}\equiv(0,\hat{\bm{z}})$ points along the direction of the external magnetic field. Exploiting orthogonality of the basis tensors in Eq.~\eqref{eq.Q}, the form factors are $\Pi_A$, $\Pi_B$, $\Pi_C$ and $\Pi_L$ of the self energy in Eq.~\eqref{eq.PiB.dec} can be extracted to be
\begin{eqnarray}
	\Pi_L&=&\frac{1}{\util^2}u_\mu u_\nu\Pi^{\mu\nu},~~
	\Pi_B=\frac{1}{\btil^2} \Big\{ b_\mu b_\nu \Pi^{\mu\nu}+\frac{\FB{b\cdot\util}^2}{\util^2}\Pi_L-2\frac{b\cdot\util}{\util^2}u_\mu b_\nu \Pi^{\mu\nu} \Big\}, \label{FF1} \\
	\Pi_A&=&\FB{g_{\mu\nu}\Pi^{\mu\nu}-\Pi_L-\Pi_B},~~
	\Pi_C=\frac{1}{\sqrt{\util^2\btil^2}} \Big\{u_\mu b_\nu\Pi^{\mu\nu}-(b\cdot\util)\Pi_L\Big\}. \label{FF2}
\end{eqnarray}
Having decomposed the self energy $\Pi_{\munu}$, the Dyson-Schwinger equation \eqref{Dyson} can now be solved for the exact propagator $D^{\mu\nu}(q)$ as~\cite{Ghosh:2019fet},
\begin{eqnarray} 
	{D}^{\mu\nu}(q;T,eB)&=&\frac{\Pcal_A^{\mu\nu}}{q^2-m_\rho^2+\Pi_A} 
	+\frac{(q^2-m_\rho^2+\Pi_L)\Pcal_B^{\mu\nu}}{\FB{q^2-m^2_\rho+\Pi_B}\FB{q^2-m_\rho^2+\Pi_L}-\Pi^2_C}  
	-\frac{\Pi_C \Pcal_C^{\mu\nu}}{\FB{q^2-m_\rho^2+\Pi_L}\FB{q^2-m_\rho^2+\Pi_B}-\Pi_C^2}  \nn \\
	&& +\frac{(q^2-m_\rho^2+\Pi_B)\Pcal_L^{\mu\nu}}{\FB{q^2-m_\rho^2+\Pi_B}\FB{q^2-m_\rho^2+\Pi_L}-\Pi^2_C}  - \frac{q^\mu q^\nu}{q^2m_\rho^2}.  \label{eq.Dbar.B}
\end{eqnarray}
The imaginary part of Eq.~\eqref{eq.Dbar.B} gives the thermomagnetic spectral function $\mathcal{S}(q;T,eB)$ as per Eq.~\eqref{eq.z.1}, and subsequently the DPR is obtained by substituting it in Eq.~\eqref{DPR}. The evaluation of $\rho^0$-meson self energy $\Pi_{\munu}(q;T,eB)$ at finite temperature in the presence of arbitrary external magnetic field has been shown in the next subsection.

\subsection{One Loop $\rho^0$-meson Self Energy in a Thermomagnetic Background}
\begin{figure}[h]
	\includegraphics[scale=.3]{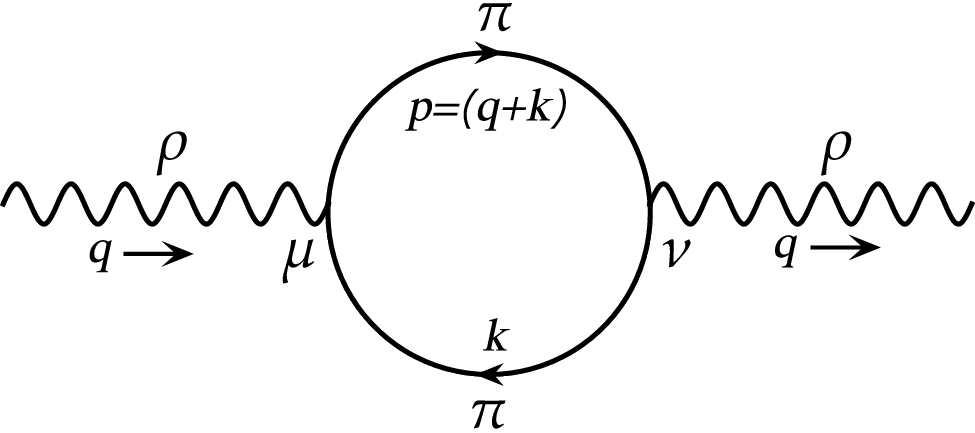}	
	\caption{Feynman diagram for one-loop self-energy of $\rho$-meson. The arrows represent the direction of momentum.}\label{FDSE}
\end{figure}

The calculation of $\rho^0$-meson thermomagnetic self energy function $\Pi_{\munu}(q;T,eB)$ and the corresponding form factors of Eqs.~\eqref{FF1} and \eqref{FF2} have been discussed in detail in Ref.~\cite{Mondal:2023vzx} and we do not repeat them here. However, we outline some of the basic steps for the sake of completeness. $\Pi_{\munu}(q;T,eB)$ will be calculated employing the effective field theory of hadrons in which the $\rho$-meson interacts with the pions via the lowest order Lagrangian (density)~\cite{Krehl:1999km}
\begin{equation}
	\mathcal{L}_\text{int}=-g_{\rho\pi\pi}~(\partial_\mu\bm{\rho}_\nu)\cdot\FB{\partial^\mu\bm{\pi}\times\partial^\nu\bm{\pi}} \label{SELint}. 
\end{equation}
in which, $\bm{\rho}_\nu$ and $\bm{\pi}$ are the iso-vector fields representing the iso-triplets of $\rho$-mesons and pions respectively, and the coupling constant $g_{\rho\pi\pi}=20.72$ GeV$^{-2}$ has been calculated from the vacuum decay width $\Gamma_{\rho\to\pi\pi}=155.8$ MeV. Owing to $\mathcal{L}_\text{int}$ in Eq.~\eqref{SELint}, the corresponding Feynman diagram for one-loop self-energy of $\rho^0$ is shown in Fig.~\ref{FDSE}.

In order to calculate the thermomagnetic self energy of $\rho^0$-meson, we employ the Real Time Formalism (RTF) of finite temperature field theory in a magnetic background. In RTF, any two-point function such as the self energy assume $2\times2$ matrix structure~\cite{Bellac:2011kqa,Mallik:2016anp}; however the estimation of 11-component of this matrix is sufficient for our purpose. The 11-component of the one-loop real time $\rho^0$-meson self energy matrix is given by (by applying the finite temperature Feynman rules to Fig.~\ref{FDSE}),
\begin{equation}
	\Pi^{\mu\nu}_{11}(q;T,eB)=i \int\! \frac{d^4k}{(2\pi)^4}\Ncal^{\mu\nu}(q,k)D^\text{Mag}_{11}(k)D^\text{Mag}_{11}(p=q+k)\label{SEPi11}
\end{equation}
where, $\Ncal^{\mu\nu}(q,k)=g^2_{\rho\pi\pi}\TB{q^4k^\mu k^\nu+(q\cdot k)^2q^\mu q^\nu-q^2(q\cdot k)\FB{q^\mu k^\nu+q^\nu k^\mu}}$ 
contains the factors coming from the interaction vertices, $D^\text{Mag}_{11}(p)$ is the 11-component of the real time charged pion propagator obtained using the Schwinger proper-time formulation (for a constant external magnetic field $\bm{B}=B\hat{\bm{z}}$ along the positive-$\hat{\bm{z}}$ direction)~\cite{Ghosh:2019fet,Ayala:2016awt} as
\begin{eqnarray}
	D^\text{Mag}_{11}(p)&=&\sum_{l=0}^{\infty}2(-1)^le^{-\alpha_p}L_l(2\alpha_p)\TB{\frac{-1}{p_\parallel^2-m^2_l+i\epsilon}+2\pi i\eta(p\cdot u)\delta(p_\parallel^2-m^2_l)}\label{DeltaPi2}
\end{eqnarray}
in which the sum is over the Landau level $l$, $\alpha_p=-p_\perp^2/eB>0$, $m_l=\sqrt{m^2_\pi+(2l+1)eB}$ is the effective pion mass in the $l^\text{th}$ Landau level, $L_l(z)$ is Laguerre polynomial of order $l$, $\eta(x)=\Theta(x)f_\text{BE}(x)+\Theta(-x)f_\text{BE}(-x)$, $p_{\parallel,\perp}^\mu = g_{\parallel,\perp}^\munu p_\nu$ with $g_\parallel^\munu=\text{diag}(1,0,0,-1)$ and $g_\perp^\munu=\text{diag}(0,-1,-1,0)$. Note that in this convention $p_\parallel^2=(p_0^2-p_z^2)$ and $p_\perp^2=-(p_x^2+p_y^2)<0$. 

Having calculated the 11-component of the real time self energy matrix, it is now easy to obtain the analytic thermomagnetic self energy function $\Pi^\munu(q;T,eB)$ that appears in the Dyson-Schwinger equation~\eqref{Dyson} via the relations $\text{Re}\Pi^{\mu\nu}(q)=\text{Re}\Pi^{\mu\nu}_{11}(q)$ and $\text{Im}\Pi^{\mu\nu}(q)=\tanh\FB{\frac{|q^0|}{2T}}\text{Im}\Pi^{\mu\nu}_{11}(q)$. Now performing the $d^2k_\parallel$ integration in Eq.~\eqref{SEPi11}, the imaginary part of the analytic self energy can be simplified to obtain the following expression
\begin{eqnarray}
	\text{Im}\Pi^{\mu\nu}(q;T,eB)&=&\tanh\FB{\frac{|q^0|}{2T}}\sum_{n=0}^{\infty}~\sum_{l=0}^{\infty}\frac{-1}{4\lambda^{1/2}(q_\parallel^2,m_l^2,m_n^2)}\sum_{k_z \in \{k_z^\pm\}} 
	\Big[ (1+f_l^k+f_n^p+2f_l^kf_n^p) \Big\{ \Ncal^{\mu\nu}_{nl}(k^0=-\omega_l^k) \Theta(q^0-E_U) \nn \\
	&& + \Ncal^{\mu\nu}_{nl}(k^0=\omega_l^k)\Theta (-q^0-E_U) \Big\}
	+(f_l^k+f_n^p+2f_l^kf_n^p) \Big\{\Ncal^{\mu\nu}_{nl}(k^0=-\omega_l^k)\Theta(q^0-E^\text{min}_L) \Theta( -q^0+E^\text{max}_L)\nn\\&& + \Ncal^{\mu\nu}_{nl}(q,k^0=\omega_l^k,k_z)\Theta(-q^0-E^\text{min}_L) \Theta( q^0+E^\text{max}_L) \Big\}
	\Big] \label{ImPi3}
\end{eqnarray}
where $\lambda(x,y,z)=x^2+y^2+z^2-2xy-2yz-2zx$ is the K\"all\'en function, $k_z^{\pm}=\frac{1}{2q_\parallel^2}\TB{-{\tilde{q}_\parallel^2}q_z\pm|q^0|\lambda^{1/2}(q_\parallel^2, m_l^2, m_n^2)}$, ~$\tilde{q}_\parallel^2=\fb{q_{\parallel}^2+m_l^2-m_n^2}$, $f^k_l=f_\text{BE}(\omega^k_l)$, $f^p_n=f_\text{BE}(\omega^p_n)$, $\omega_l^k=\sqrt{k_z^2+m_l^2}$, $\omega_n^p=\sqrt{p_z^2+m_n^2}$, $E_U=\sqrt{q_z^2+\FB{m_l+m_n}^2}$, $E^\text{(min,max)}_L = \text{(min,max)}(q_z, E^{\pm})$, $E^{\pm}=\frac{m_l-m_n}{|m_l\pm m_n|}\sqrt{q_z^2+\FB{m_l\pm m_n}^2}$, and  
\begin{equation}\label{Eq.Numununl}
	\Ncal^{\mu\nu}_{nl}(q,k_\parallel)=4(-1)^{n+l}\int\!\frac{d^2k_\perp}{\fb{2\pi}^2}e^{-\alpha_k-\alpha_p}\Ncal^{\mu\nu}(q,k)L_l(2\alpha_k)L_n(2\alpha_p).
\end{equation}

In Eq.~\eqref{ImPi3}, there appears a number of $\Theta$-functions which actually represent branch cuts of the self energy function in the complex energy plane of the $\rho_0$-meson and it physically correspond to kinematically allowed scattering and decay processes involving charged pions occupying different Landau levels. In particular $E_U$ and $E^\text{(min,max)}_L$ in Eq.~\eqref{ImPi3} respectively denote the Unitary and Landau cut thresholds for a particular $(l,n)$.


\section{Numerical Results}\label{Numerical}
In this section we have performed numerical analysis of DPR from a magnetized hadronic matter as function of different kinematic variables, such as, the invariant mass $ M $, the transverse momentum $ q_T $ with respect to beam axis  and the azimuthal angle $ \phi $ (see Fig.~\ref{Frame}). We present results for central rapidity which corresponds to $ q_x = 0 $. Since, we choose the constant  background magnetic field along $z$-axis, it will  break the rotational symmetry in $ yz $-plane which is also the perpendicular plane with respect to the beam axis. So it is expected that the DPR should be anisotropic in the transverse direction. To quantify this, we have also examined numerically the ellipticity of the dilepton production  by studying the flow co-efficient $ v_2 $ at mid-rapidity as a function of $ M $ for different values of $ q_T $. Since we are interested in DPR for hadronic medium, we have considered two different values of medium temperature, $ T = 130 $ and $ 160 $ MeV respectively, representing the hadronic stage of the matter created in HICs. As the hadrons are created in the later stages of HICs, it is expected that the strength of the background field should be smaller compared to the QGP phase. So we will choose two representative values, $ eB = 0.02 $ and $ eB = 0.05 $ GeV$ ^2 $ which will provide us the opportunity to explore the interplay between the magnetic field and the thermal effects. In this section, all the numerical calculations  with finite values of background magnetic field (where the sum over infinite numbers of Landau levels has to be performed, for example see Eq. \eqref{ImPi3}) are evaluated by considering upto 500 Landau levels ensuring the convergence of the results. 

At first we will study the total differential rate by integrating over the whole range of the azimuthal angle. This can be achieved by performing the following integral:
\begin{equation}
	\frac{dN}{d^{4}xMdMq_Tdq_Tdy}=\int_{0}^{2\pi}d\phi\frac{dN}{d^4xd^4q}\label{DTDPR}
\end{equation}
where $ {dN}/{d^4x d^4q} $ is given by Eq.~\eqref{DPR}. The numerical evaluation of this integral, which involves summing over a large number Landau levels, can be simplified employing spatial symmetries of the magnetized matter at mid rapidity~\cite{Wang:2022jxx}. Firstly, we note that, with our choice of axes, the system is invariant under spatial rotation about $ z $ axis, which is a subgroup of the spatial rotation group that remains unbroken in $ B\ne 0 $ case. Invoking this symmetry we can write $\frac{dN}{d^4xd^4q}(\pi-\phi)=\frac{dN}{d^4xd^4q}(\phi) $. Again, since the magnetic field is an axial vector, there is also a reflection symmetry with respect to the reaction plane. This will allow us to write $\frac{dN}{d^4xd^4q}(\phi) = \frac{dN}{d^4xd^4q}(-\phi)$. Hence, we need to evaluate the integral only in the first quadrant, i.e. $ 0 \le \phi  < \pi/2 $. The results for the other three quadrants can be evaluated using the above symmetries.
\begin{figure}[h]	
 	\includegraphics[angle = -90, scale=0.35]{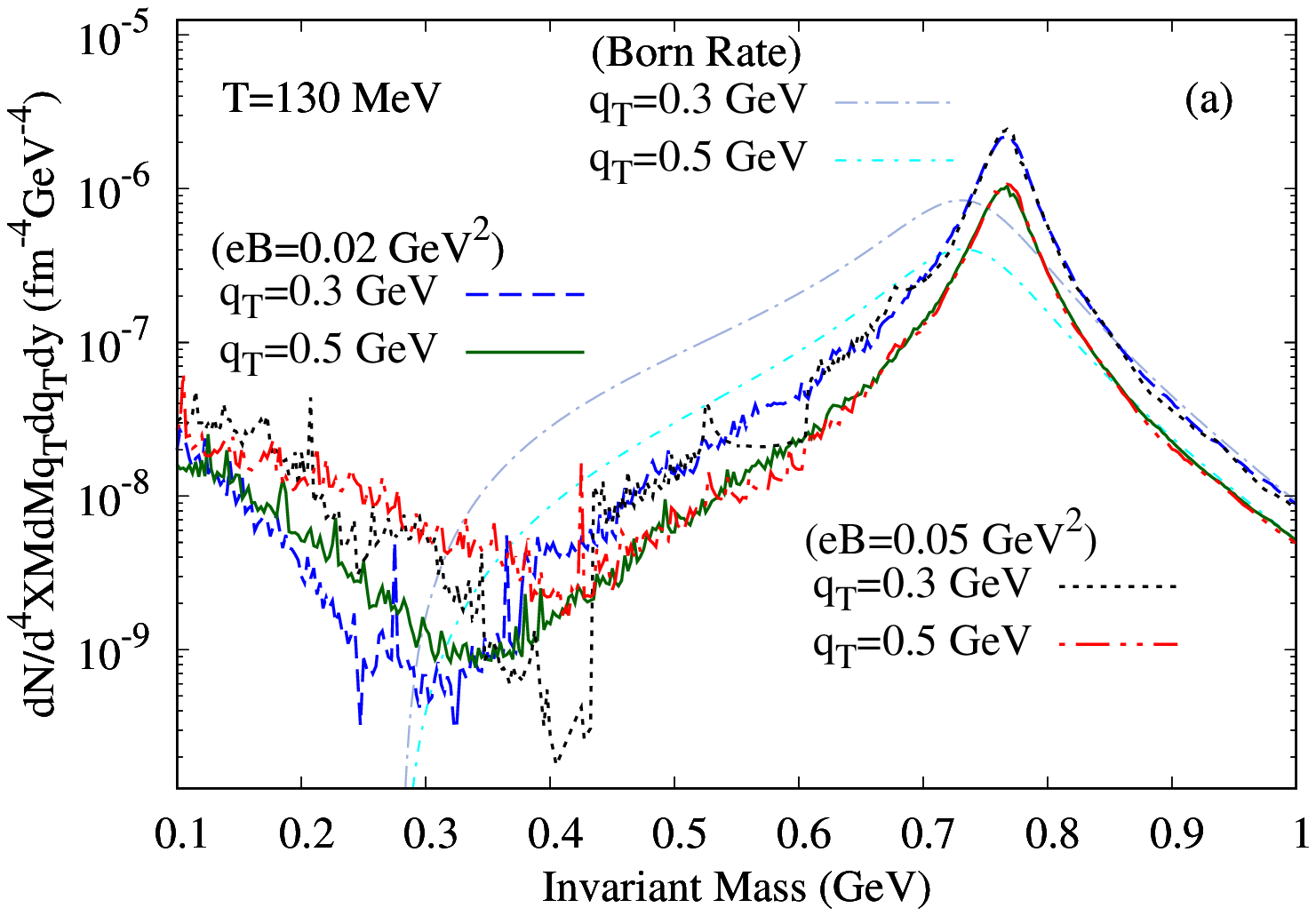}
 	\includegraphics[angle = -90, scale=0.35]{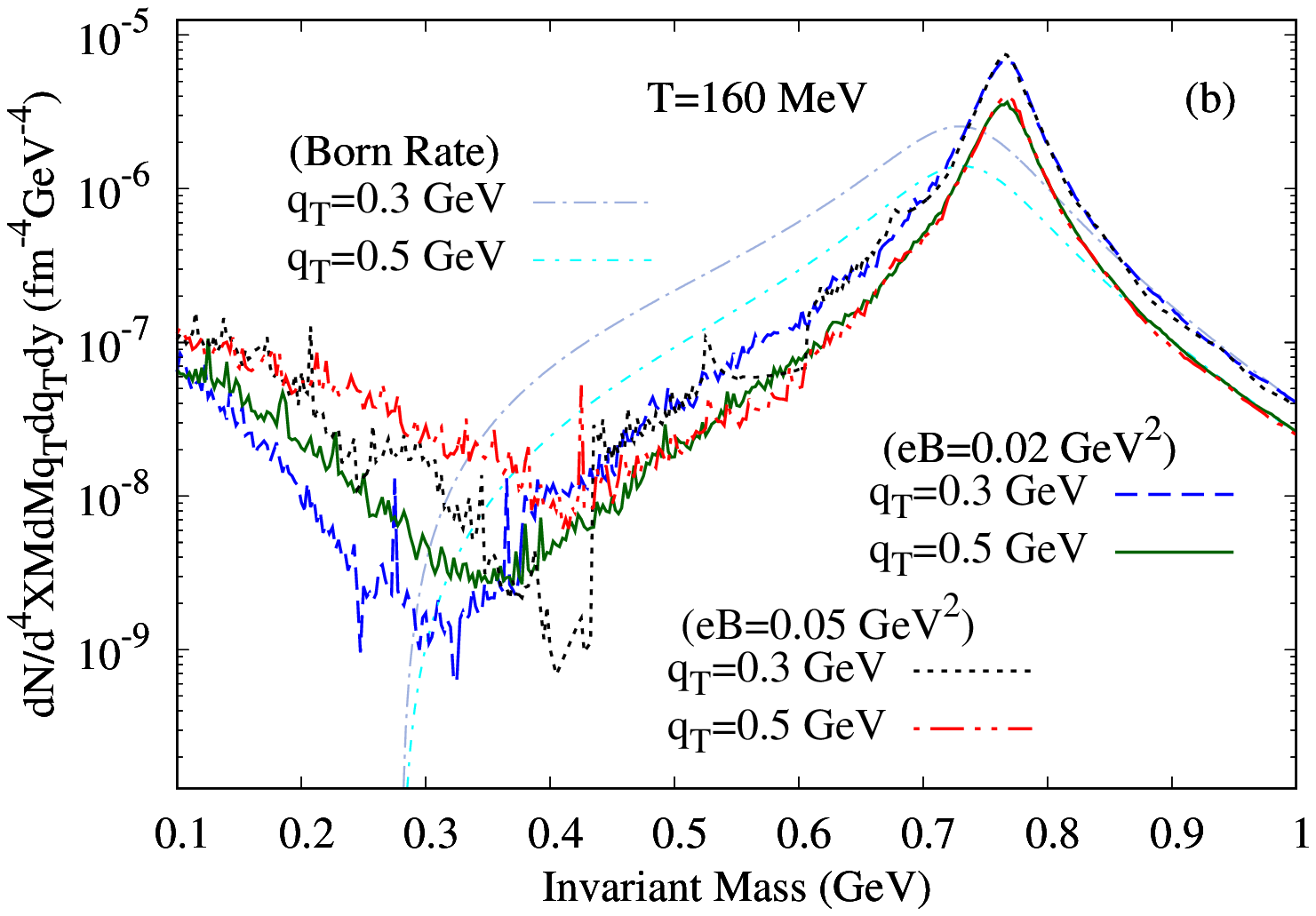}
 	\caption{(Color online) Total differential rate as a function of invariant mass for $eB=0.02,~0.05 \rm~GeV^2$ and $q_T=0.3,~0.5~\rm GeV$ at (a)~$T=130\rm~MeV$ and (b) $T=160~\rm~MeV$. The born rate in absence of the background field is also shown for comparison; gray and cyan lines represent born rate corresponding to $q_T=0.3$ and $0.5$ GeV respectively.}
 	\label{DPRM}
\end{figure}

 In Figs.~\ref{DPRM}(a) and (b), we have shown the variation of the total differential rate as a function of invariant mass for different values of $ eB $ and $ q_T $ at $ T=130 $ and $ T=160 $ MeV respectively. For comparison, we have also shown the born rate in absence of the background field. From both the plots, it is evident that the rate is highly enhanced in the low invariant mass region compared to the born rate in absence of $ eB $ owing to  the Landau cut contributions. This is a purely magnetic field dependent effect and can be attributed to the process where a $\rho$-meson is absorbed by means of scattering with a pion in lower Landau level producing a pion in higher Landau level in the final state  (and the time reversed process). For intermediate values of invariant mass, the total differential rate is smaller compared to the zero field case due to the shift of unitary cut contributions towards higher values of invariant mass. This can be seen explicitly if we work in the limit $ q_\perp = 0 $, as studied in~\cite{Mondal:2023vzx}, where the Unitary and Landau cut contributions are separated from each other. However, in the present case, the Landau cut contribution remains finite for whole range of $ M $, giving rise to the specific behaviour seen in Figs.~\ref{DPRM}(a) and (b).  The higher invariant mass region is dominated by the Unitary cut contributions which corresponds to the decay $ \rho^0\rightarrow \pi^+\pi^- $. Here the effect of the background field is not that significant as compared to that in the lower invariant mass region and the plots for $ eB = 0.02 $ and $ 0.05 $ GeV$ ^2 $ for different values of $ q_T $ follow the $ eB = 0 $ results. Furthermore, it can be observed that for higher values of temperature the overall magnitude of the DPR increases, as noticeable by  comparing of Figs.~\ref{DPRM}(a) and (b), which is a consequence of the enhancement in the availability of the thermal phase space. Finally, for a given value of $ eB $, the dileptons with high $ q_T $ are thermally suppressed as can be seen in both the figures, which will be more evident from the next paragraph.   
  Few comments on the spikelike structures seen in both the plots are in order here. This nonsmooth $ M $-dependence of the total differential rate correspond to the so called threshold singularities which appear whenever the energy of the virtual photon becomes equal to one of the numerous Landau-level thresholds~\cite{Wang:2021ebh}. This results in an abrupt change in kinematically allowed phase space for dilepton production (see \cite{Ghosh:2018xhh,Mondal:2023vzx} for detailed description of the analytic structure of $ \IM\Pi^\munu $ in the complex $ q^0 $-plane). Mathematically, this is related to the scenario when the K\"all\'en function appearing in the denominator of Eq.~\eqref{ImPi3} becomes vanishingly small.
\begin{figure}[h]	
	\includegraphics[angle = -90, scale=0.35]{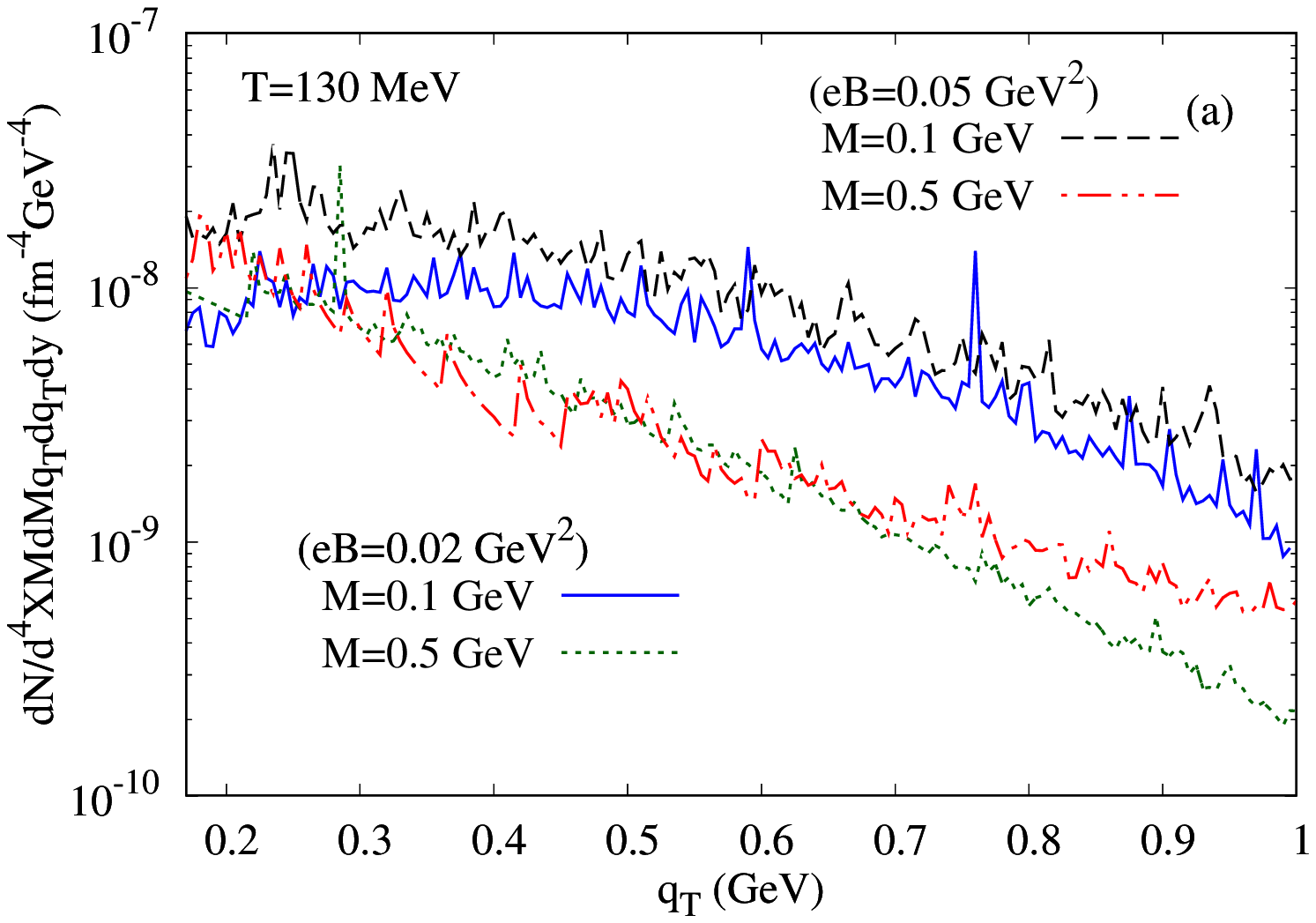}
	\includegraphics[angle = -90, scale=0.35]{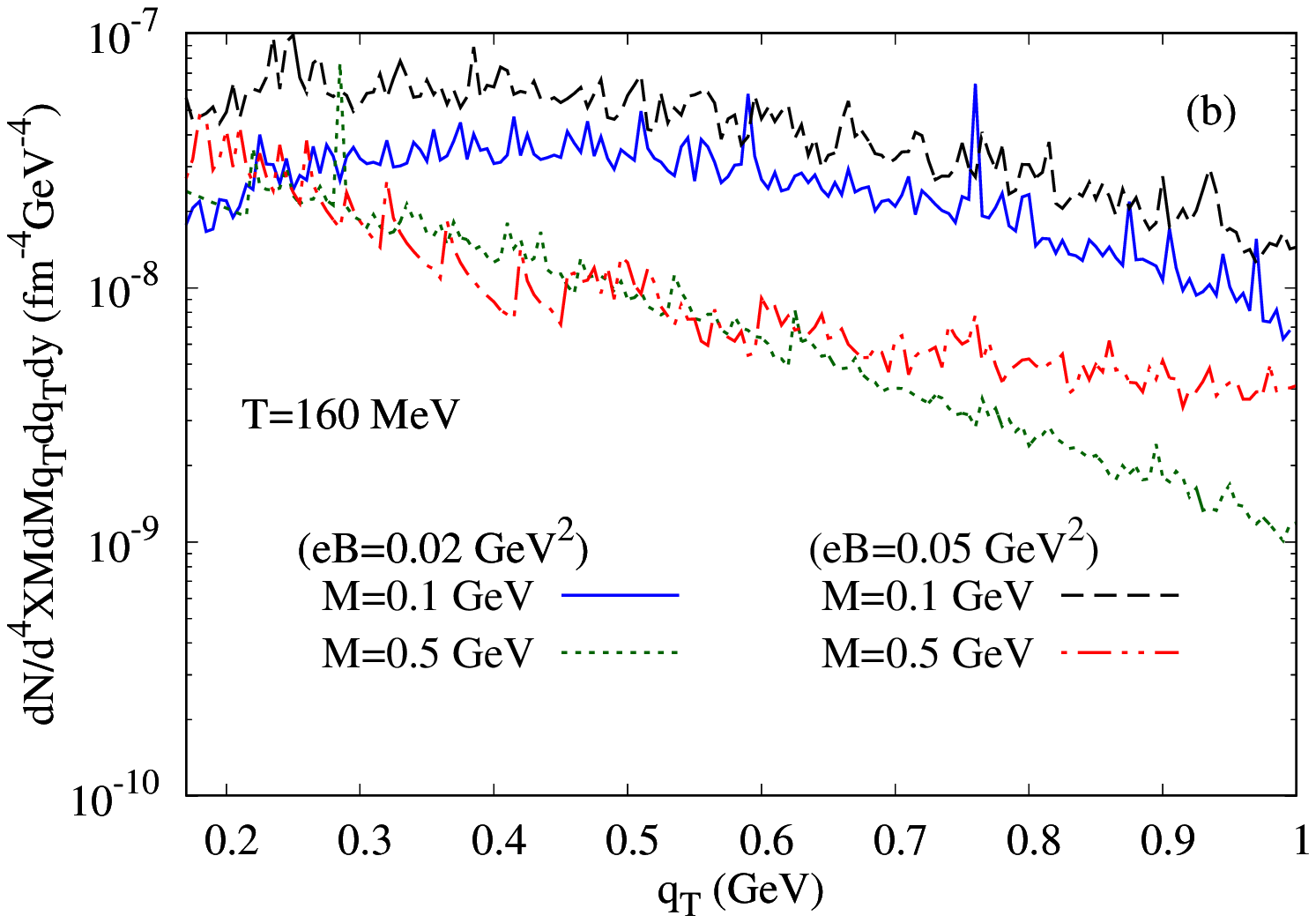}
	\caption{(Color online) The variantion of total differential rate with $q_T$ for $eB=0.02,~0.05 \rm~GeV^2$ and $M=0.1,~0.5~\rm GeV$ at (a)~$T=130\rm~MeV$ and (b) $T=160~\rm~MeV$.}
	\label{DPRMqT}
\end{figure}  

In Figs.~\ref{DPRMqT}(a) and (b), the $ q_T $ dependence of the total differential rate is presented for several values of $ eB $ and invariant mass for $ T=130 $ and $ 160  $ MeV respectively. As observed earlier, the overall rate is a decreasing function of the transverse momentum owing to the thermal suppression of high energy dileptons which is evident from both the plots. The origin of this suppression are two fold. Primarily this comes from the overall Bose distribution function present in the expression of DPR given by Eq.~\eqref{DPR}. The additional suppression can be understood from the expression of spectral function or equivalently imaginary part of $\rho^0$ self energy given by Eq.~\eqref{ImPi3} which also contains the Bose distribution functions.  Moreover, in both Figs.~\ref{DPRMqT}(a) and (b), for lower values of the invariant mass i.e. at $M=0.1 $ GeV, the overall magnitude of the total differential rate is larger for higher value of $ eB $ for the whole range of $ q_T $ owing to the enhancement in the Landau cut contribution which is the only kinematically allowed contributions~\cite{Mondal:2023vzx} (see the solid and dash-dot line). However, at $ M= 0.5 $ GeV, Landau cut contribution decreases and unitary cut contribution starts to develop. As a result, no generic background field dependence is observed. Furthermore, for higher values of $ T $ the observed increase in the overall magnitude of the total differential rate is due to the enhancement in the available thermal phase space, a result which is also evident in Figs.~\ref{DPRM}(a) and (b). It is well known that the inverse slope of the transverse momentum distribution is a measure of the effective temperature of the medium. It can be inferred from Figs.~4 (a) and (b) that the presence of magnetic field can result in modification of this slope indicating  a change in effective temperature of the medium. However, to get a meaningful estimate one has to perform full space-time evolution of this rate using magneto-hydrodynamics which is beyond the scope of the present work.

%
\begin{figure}[h]	
	\includegraphics[angle = -90, scale=0.50]{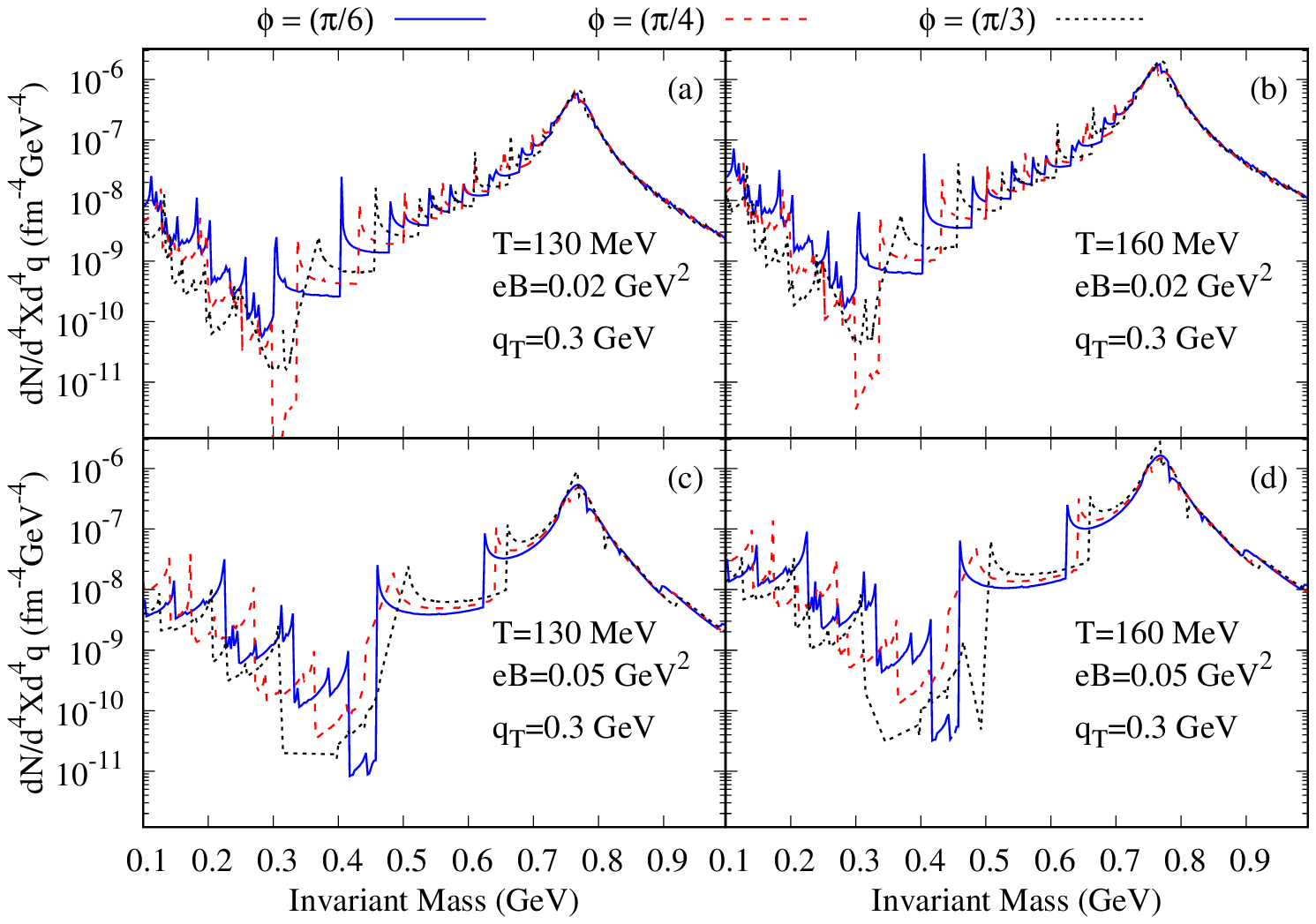}
	\caption{(Color online) DPR as a function of $M$ for $q_T=0.3~\rm~GeV$ and different values of $\phi$ at (a) $T=130$ MeV, $eB=0.02~\rm~GeV^2$, (b) $T=160$ MeV,  $eB=0.02~\rm~GeV^2$, (c) $T=130$ MeV, $eB=0.05~\rm~GeV^2$, and, (d) $T=160$ MeV, $eB=0.05~\rm~GeV^2$.}
	\label{MDPRMqTFi1}
\end{figure}
\begin{figure}[h]	
	\includegraphics[angle = -90, scale=0.50]{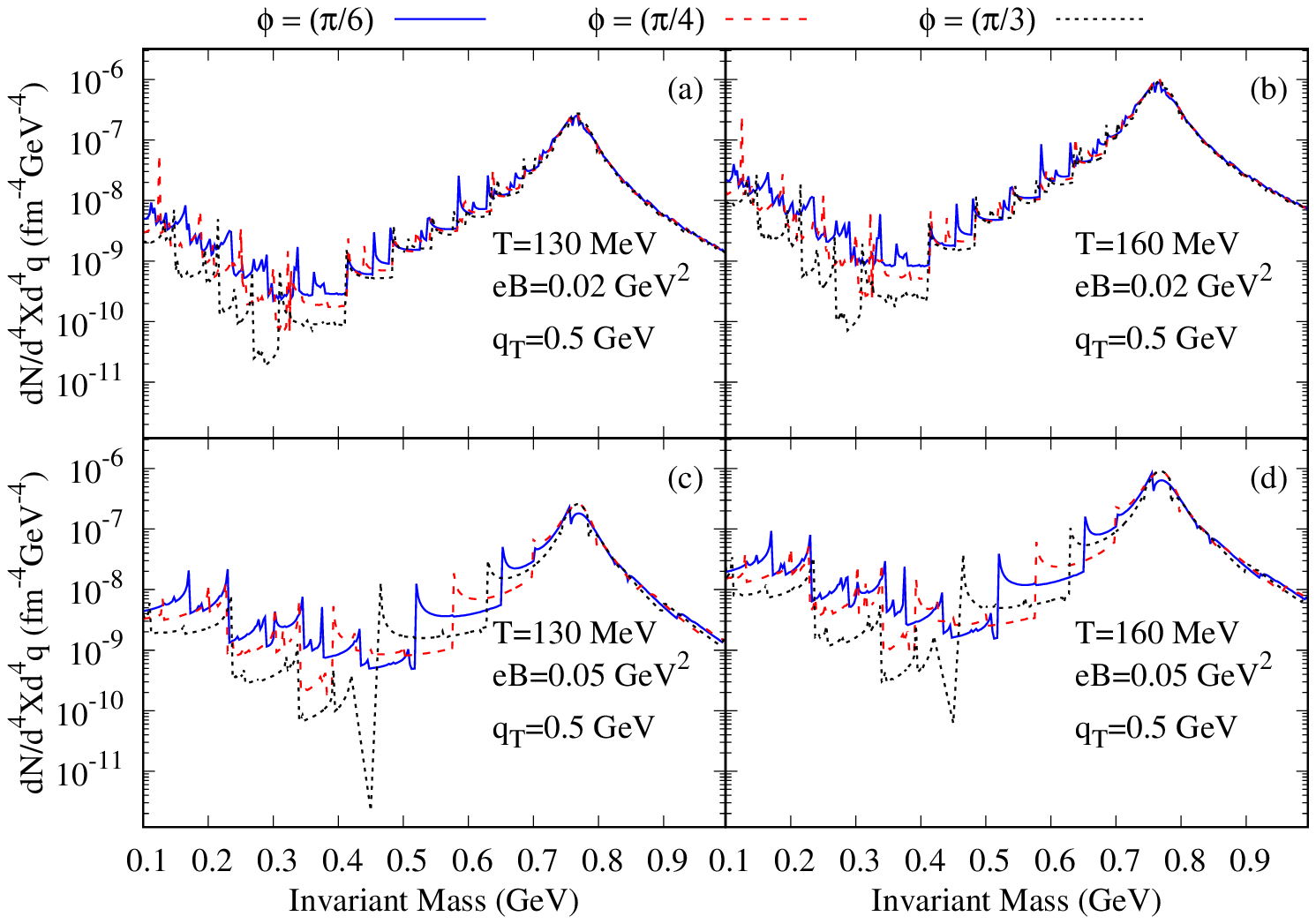}
	\caption{(Color online) DPR as a function of $M$ for $q_T=0.5~\rm~GeV$ and different values of $\phi$ at (a) $T=130$ MeV, $eB=0.02~\rm~GeV^2$, (b) $T=160$ MeV,  $eB=0.02~\rm~GeV^2$, (c) $T=130$ MeV, $eB=0.05~\rm~GeV^2$, and, (d) $T=160$ MeV, $eB=0.05~\rm~GeV^2$.}
	\label{MDPRMqTFi2}
\end{figure}

Now, to study the anisotropic nature of the DPR, it is useful to examine the angular dependence of the rate. We have shown this in Figs.~\ref{MDPRMqTFi1} and \ref{MDPRMqTFi2}. Each figure contains a set of four panels in which $ T $ increases in the horizontal direction and $ eB $ increases in the downward direction. As already discussed, considering the symmetry arguments, the knowledge of DPR  in the first quadrant is sufficient to evaluate the results in the other three quadrants. So in all the plots we have chosen  $\phi=\frac{\pi}{3}$, $\frac{\pi}{4}$, and $~\frac{\pi}{6}$ as representative values of the azimuthal angle. In Figs.~\ref{MDPRMqTFi1}(a)-(d), we have depicted variation of DPR as a function of invariant mass at $ q_T = 0.3 $ GeV in three different azimuthal direction as indicated before for different values of $T$ and $eB$. It is evident that DPR has a non-trivial $ \phi $-dependence in the lower values of invariant mass indicating the anisotropic nature of DPR in the transverse plane with respect to the beam axis. This anisotropy qualitatively increases as we increase the background field strength as understandable by comparing figures along the vertical direction.  However, the nonsmooth nature of the spectrum does not lead to any generic trend in $ \phi $-dependence of the DPR. On the other hand, for high values of invariant mass, the rate has a negligible $\phi $-dependence in all the cases indicating an isotropic emission of high mass dileptons. Moreover, as we move along the horizontal panels, the overall magnitude of the DPR increases owing to the increase in thermal phase space as indicated earlier.  In Figs.~\ref{MDPRMqTFi2} (a)$ - $(d), we have studied the $ \phi $-dependence of the DPR at $ q_T  = 0.5$ GeV keeping all the other physical parameters fixed. The qualitative nature of the plots remains similar with the previous observations except the yield is slightly lower. It should be noted that, spikelike structure appearing in Figs.~\ref{MDPRMqTFi1} and \ref{MDPRMqTFi2} are the consequence of Landau level threshold effect as discussed earlier.
\begin{figure}[h]	
	\includegraphics[angle = -90, scale=0.5]{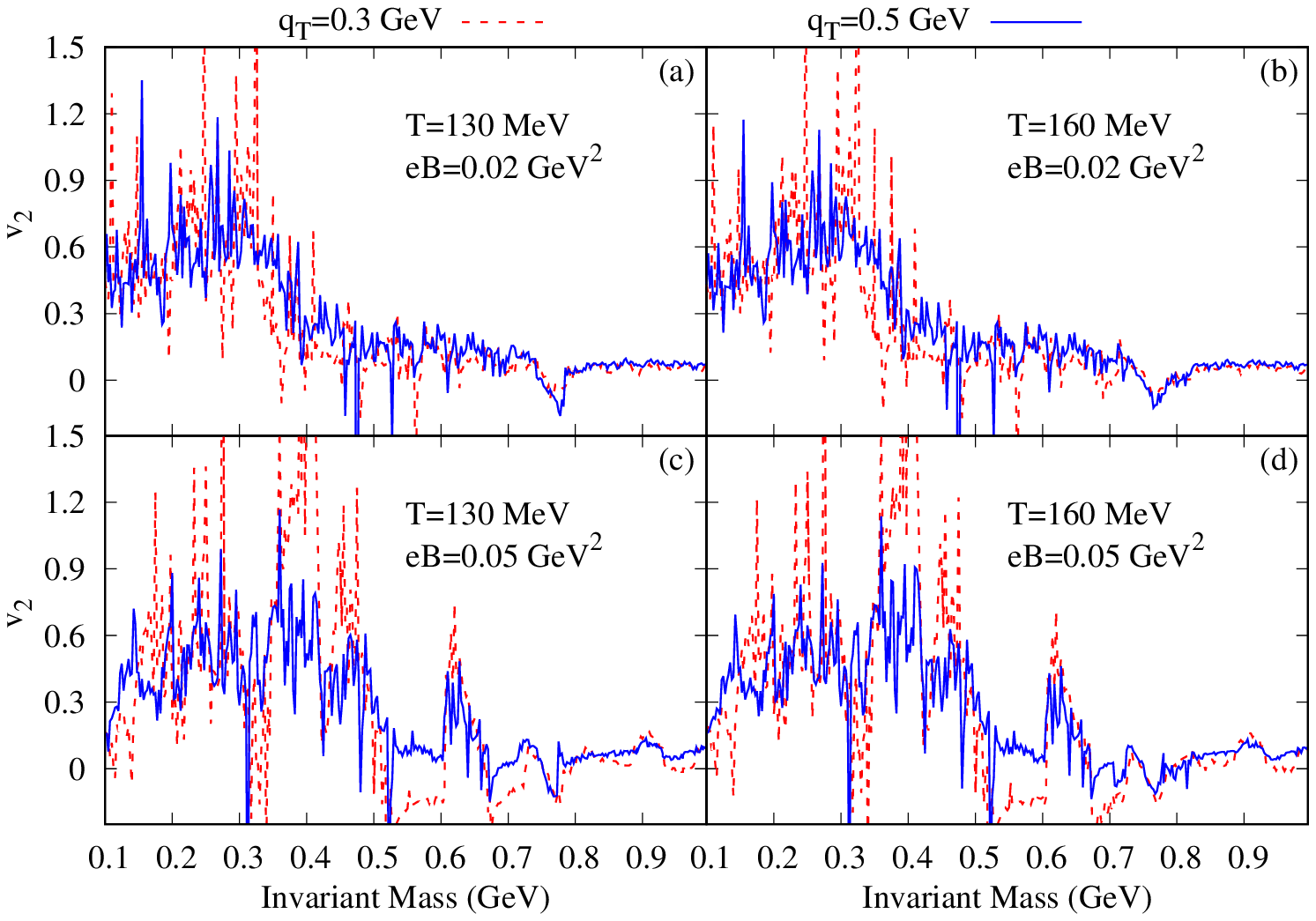}
	\caption{(Color online) Ellipiticity of dilepton production $v_2$ as a function of invariant mass $M$ for $q_T=0.3$ and $0.5$ GeV at (a) $T=130$ MeV, $eB=0.02~\rm~GeV^2$, (b)~$T=160~\rm~MeV$,~$eB=0.02~\rm~GeV^2$ (c)$~T=130~\rm~MeV$,~$eB=0.05~\rm~GeV^2$, (d)$~T=160~\rm~MeV$,~$eB=0.05~\rm~GeV^2$}
	\label{MDPRM}
\end{figure}
\begin{figure}[h]	
	\includegraphics[angle = -90, scale=0.5]{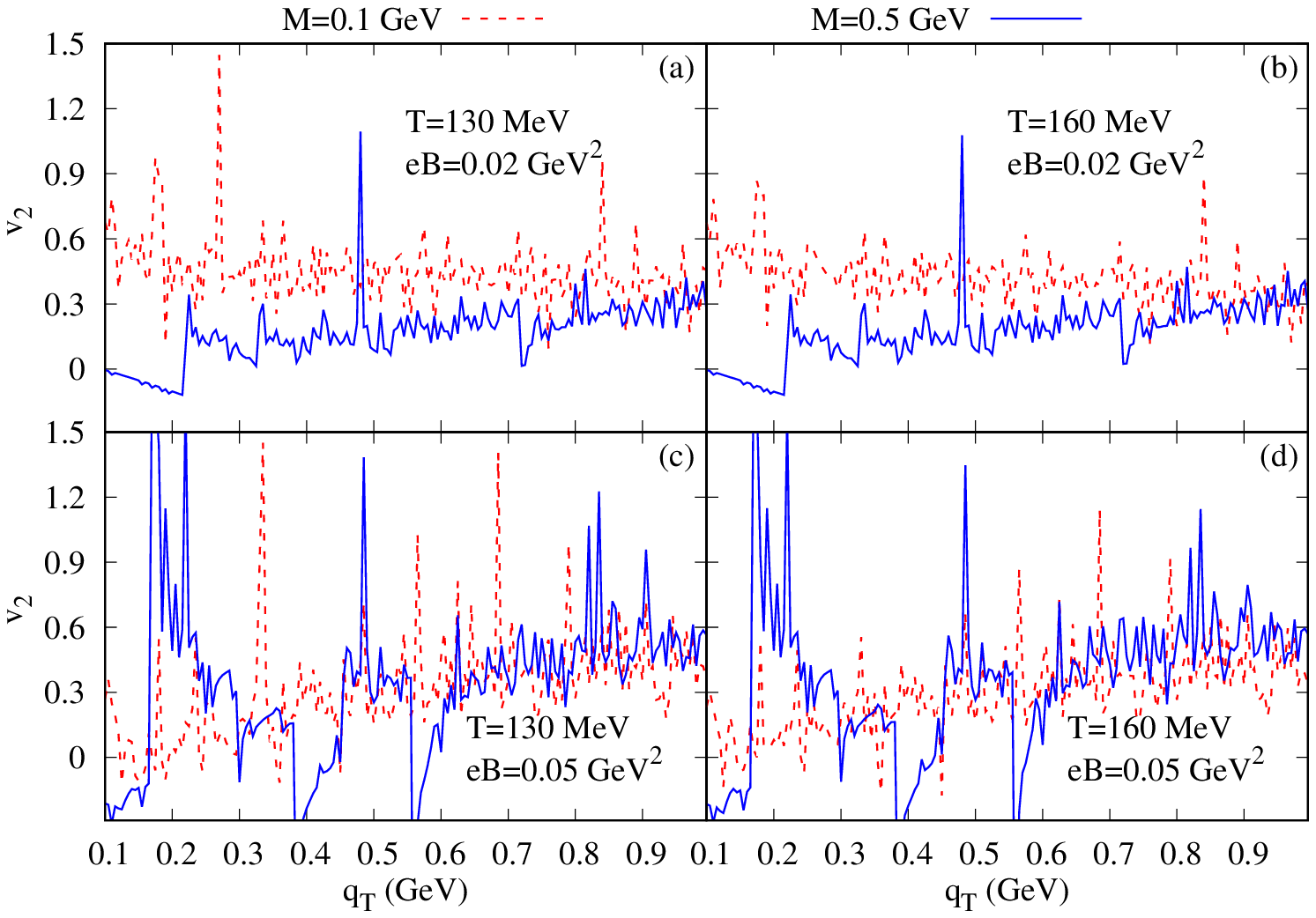}
	\caption{(Color online) Ellipiticity of dilepton production $v_2$ as a function of transverse momentum $q_T$ for $M=0.1$ and $0.5$ GeV at (a) $T=130$ MeV, $eB=0.02~\rm~GeV^2$, (b)~$T=160~\rm~MeV$,~$eB=0.02~\rm~GeV^2$ (c)$~T=130~\rm~MeV$,~$eB=0.05~\rm~GeV^2$, (d)$~T=160~\rm~MeV$,~$eB=0.05~\rm~GeV^2$}
	\label{qTV2}
\end{figure}

As mentioned earlier, the amount of anisotropy in dilepton production can be quantified by studying the conventional elliptic flow parameter $ v_2 $ defined in Eq.~\eqref{Vn}. In Figs.~\ref{MDPRM}(a)-(d), we have presented ellipticity parameter $ v_2 $ as a function of $ M $ for different values of transverse momenta in the similar setting namely in the horizontal (downward) direction the temperature (background magnetic field) increases keeping $ eB  $ $ (T) $ constant. As evident from all the figures, $ v_2 $ is also affected by the appearance of several spikelike structures corresponding to the numerous Landau level thresholds. However, many qualitative features can be identified from the numerical data. Notice that, there exists a strong tendency of $ v_2 $ to remain positive in low invariant mass region for small $ eB $-values. This behaviour is further enhanced at large $ q_T $. These observations are consistent with the angular dependence of the DPR shown in Figs.~\ref{MDPRMqTFi1} and \ref{MDPRMqTFi2}. The positive values of $ v_2 $ imply an oblate shape of the DPR and indicates the production is larger along the direction transverse to the background field i.e. $ \phi \sim 0$ which corresponds to the reaction plane. However, a prediction of overall ellipticity is restricted as $ v_2 $, although generally non vanishing, may change sign in the lower invariant mass specifically for small $ q_T $-value. This is evident in case of stronger magnetic field presented in the lower panels. Moreover, in all the cases, for high values of invariant mass $ v_2 $ shows highly oscillatory behaviour around zero implying a isotropic nature of the differential rate which can be confirmed from the earlier observations made in Figs~\ref{MDPRMqTFi1} and \ref{MDPRMqTFi2}. The qualitative behaviour of $ v_2 $ noticed in Figs.~\ref{MDPRM}(a)-(d), is consistent with the earlier observations made in~\cite{Wang:2022jxx} related to dilepton emissions from a magnetized quark-gluon plasma. 

Next we plot the ellipticity parameter $v_2$ of dilepton emission rate as a function of transverse momentum $q_T$ in Figs.~\ref{qTV2}(a)-(d) for different values of invariant mass $M=0.1, 0.5$ GeV, temperatures $T=130, 160~\rm MeV$ and magnetic field $eB=0.02, 0.05~\rm GeV^2$. We see that $v_2$ does not show any strong dependence on temperature. But, there is a tendency of overall increase of $v_2$ with $q_T$. Moreover, $v_2$ is a non smooth function of $q_T$ due to the the Landau level quantization and numerous threshold effects.

\section{Summary \& Conclusion}\label{SC}
In this work we have studied the dilepton production rate from magnetized hadronic matter as a function of invariant mass $M$, transverse momentum $q_T$ (with respect to beam axis) and the azimuthal angle $\phi$ for different values of temperature and magnetic field. The primary ingredient in the emission rate of dileptons is the  thermomagnetic in-medium spectral function of the $\rho^0$-meson which has been evaluated employing the real time method of thermal field theory and Schwinger proper-time formulation. 

The total differential rate is found to be highly enhanced in the low invariant mass region due to the emergence of non-trivial Landau cut contributions in the physical kinematic region. This is an outcome of the presence of the finite background field. The appearance of such a contribution physically corresponds to the fact that while scattering with a $ \rho^0 $ meson, charged pions can occupy different Landau levels. However, at higher values of invariant mass, which is dominated by the Unitary cut contributions coming from two-pion threshold, the effect of background  field is negligible. The Landau cut contribution in the low invariant mass region is also responsible for an overall increase in the transverse momentum distribution of the dielptons with the increase of the background field. The DPR yield in the low invariant mass region is found to be highly anisotropic in different azimuthal directions. However, for high values of invariant mass, the rate has a negligible $\phi $-dependence implying an isotropic emission of high mass dileptons. These observations are verified further by examining the elliptic flow parameter  $ v_2 $. We have found that  $ v_2 $ tends to remain positive  in low invariant mass region particularly for small $ eB $-values implying the production is larger along the direction transverse to the background field.

\section*{Acknowledgments}
R.M., N.C., S.S. and P.R. are funded by the Department of Atomic Energy (DAE), Government of India. S.G. is funded by the Department of Higher Education, Government of West Bengal, India. 

\bibliographystyle{apsrev4-1}
\bibliography{Ref}

\end{document}